\documentclass[sigconf]{acmart}

\usepackage[linesnumbered,ruled,vlined]{algorithm2e}
        
\usepackage{amsmath,amssymb,amsfonts}
\usepackage{algorithmic}
\usepackage{graphicx}
\usepackage{textcomp}
\usepackage{xcolor}
\usepackage{multirow}
\usepackage{stackengine}
\usepackage{url}
\usepackage{bm}

\usepackage{arydshln}
\usepackage{flushend}

\def\BibTeX{{\rm B\kern-.05em{\sc i\kern-.025em b}\kern-.08em
    T\kern-.1667em\lower.7ex\hbox{E}\kern-.125emX}}
\usepackage{enumitem}

\definecolor{brickred}{rgb}{0.8, 0.25, 0.33}

\usepackage{xspace}
\usepackage{amsthm}

\theoremstyle{definition}

\newtheorem{lemma}{Lemma}[section]
\newtheorem{corollary}{Corollary}[section]

\newcommand{\tool}{CompVQC\xspace}
\newcommand{\framework}{CompVQC\xspace}

\usepackage{pifont}

\usepackage{hyperref}
\hypersetup{
    colorlinks=true,
    linkcolor=blue,
    citecolor=blue,
    filecolor=magenta,      
    urlcolor=cyan,
    pdftitle={Overleaf Example},
    pdfpagemode=FullScreen,
    }
\begin{document}

\title{Quantum Neural Network Compression}

\author{
Zhirui Hu$^{1,2}$, Peiyan Dong$^{3}$, Zhepeng Wang$^{1,2}$, Youzuo Lin$^{4}$, Yanzhi Wang$^{3}$, Weiwen Jiang$^{1,2}$
\\\vspace{5pt}{\large $^{1}$Electrical and Computer Engineering Department, George Mason University, Fairfax, Virginia 22030, United States \\ $^{2}$Quantum Science and Engineering Center, George Mason University, Fairfax, Virginia 22030, United States \\ $^{3}$Department of Electrical and Computer Engineering, Northeastern University, Boston, MA 02115, United States\\$^{4}$Earth and Environmental Sciences Division, Los Alamos National Laboratory, NM, 87545, USA}
\\ 
{\large (zhu2@gmu.edu; wjiang8@gmu.edu)}
}

\begin{abstract}
Model compression, such as pruning and quantization, has been widely applied to optimize neural networks on resource-limited classical devices. 
Recently, there are growing interest in variational quantum circuits (VQC), that is, a type of neural network on quantum computers (a.k.a., quantum neural networks).
It is well known that the near-term quantum devices have high noise and limited resources (i.e., quantum bits, qubits);
yet, how to compress quantum neural networks has not been thoroughly studied. 
One might think it is straightforward to apply the classical compression techniques to quantum scenarios.
However, this paper reveals that there exist differences between the compression of quantum and classical neural networks.
Based on our observations, we claim that the compilation/traspilation has to be involved in the compression process.
On top of this, we propose the very first systematical framework, namely \tool, to compress quantum neural networks (QNNs).
In \tool, the key component is a novel compression algorithm, which is based on the alternating direction method of multipliers (ADMM) approach.
Experiments demonstrate the advantage of the CompVQC, reducing the circuit depth (almost over $2.5\times$) with a negligible accuracy drop ($<$1\%), which outperforms other competitors.
Another promising truth is our CompVQC can indeed promote the robustness of the QNN on the near-term noisy quantum devices.
\end{abstract}




\begin{teaserfigure}
  \includegraphics[width=1\linewidth]{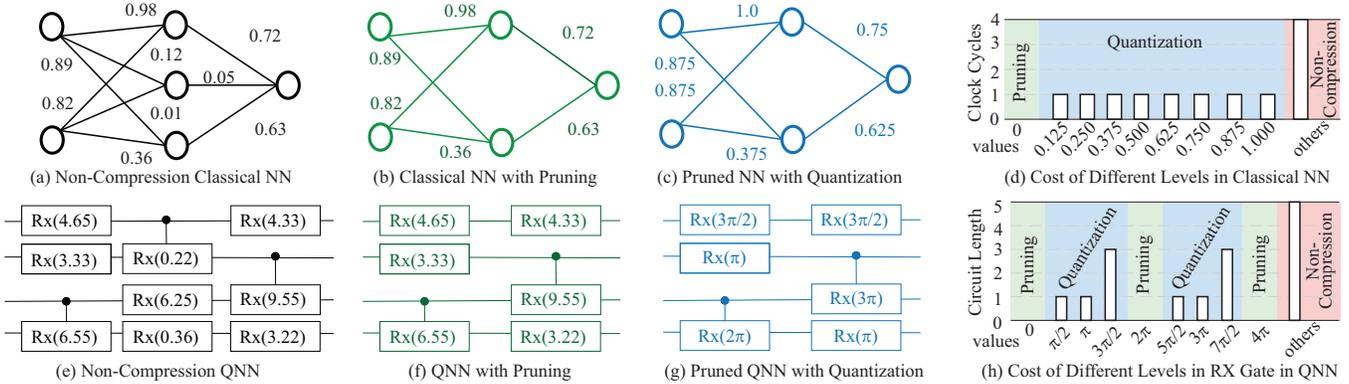}
  \caption{Compression on classical neural network and quantum neural network are fundamentally different: (a)-(c) the pruning and quantization on classical neural networks; (d) value of parameters for pruning and quantization and corresponding clock cycles/latency; (e)-(g) the pruning and quantization on variational quantum circuit, a type of quantum neural networks; (h) quantum gates can be pruned by different parameter values and different quantization values lead to different costs.}
  \label{fig:cvsq}
\end{teaserfigure}

\renewcommand{\shortauthors}{Zhirui Hu, et al.}


\maketitle

\section{Introduction}
Machine learning, in particular neural network, has shown its significant superiority on many practical tasks, like image classification, video segmentation, pattern recognition, etc. 
These tasks are not only the basis of processing classical data, like digital image or video in computer vision, texts or audio in natural language processing, but they are also able to be applied to study properties of physical systems, such as quantum phase recognition (QPR) \cite{cong2019quantum,jiang2021machine}.
Moreover, with the consistent development of quantum sensing \cite{degen2017quantum} and quantum memory \cite{lvovsky2009optical,phalak2022optimization}, data will be acquired and stored in quantum format.
This calls for the in-suit process of data in quantum computing.
All these progresses significantly push forward the study of neural networks in quantum computing.
As one of the typical quantum neural network (QNN), variational quantum circuit (VQC) is widely used in different applications \cite{ezawa2021variational,chen2020variational,khairy2020learning,wille2021introduction,ravi2021threat,burgholzer2022characteristics}.


Existing works have theoretically shown the effectiveness and efficiency of VQC in the perfect scenario \cite{bravo2020variational}; however, it is well known that the near-term quantum devices have high noise.
Coherent and incoherent errors in quantum devices will in turn limit the quantum circuit depth/length (i.e., longer circuit has larger accumulated errors)\cite{clemens2004quantum,burgholzer2022characteristics}.
As such, when it comes to deploying QNN onto the near-term quantum devices, it is critical to control the circuit depth.
In classical neural network (as shown in Figure \ref{fig:cvsq}(a)), the compression techniques (including pruning as shown in Figure \ref{fig:cvsq}(b) and quantization as shown in Figure \ref{fig:cvsq}(c)) are widely used to reduce the execution length.
Inspired by this, in this work, we explore how to compress QNN to reduce its circuit length.

It seems straightforward to apply the classical compression techniques to QNN \cite{wang2022chip,wang2021roqnn}; however, the fundamental difference on neural network designs on quantum computing bring new challenges, where QNN is composed of parameterized quantum gates, as shown in Figure \ref{fig:cvsq}(e).
First of all, it is unclear which quantum gates can be pruned without the change of the function; as shown in Figure \ref{fig:cvsq}(h), in addition to the gates with the value of $0$, there exist gates with other parameters, say $2\pi$ and $4\pi$, which can also be pruned, called ``pruning levels'' in this paper.
Second, unlike classical neural network having the same cost, the circuit depths of different parameter values in QNN are not fixed, called ``quantization levels''; as shown in Figure \ref{fig:cvsq}(h), quantization levels of $\pi/2$ and $3\pi/2$ can reduce the circuit depth from 5 to 1 and 3, respectively.
Last but not least, the ``compression levels'' (i.e., pruning level and quantization level) will not only be affected by the value of the parameter, it will also be affected by the type of gates, the quantum devices, and the compiler.

In response to all the above challenges, this paper presents a holistic framework, namely \framework, to automatically compress a given QNN.
More specifically, \framework provides the fundamental understandings of pruning and quantization in quantum settings; on top of this, a compression-level look-up-table (LUT) for quantum gates is used in the given QNN for a specific quantum backend and compiler.
With the help of compression-level LUT, \framework applies alternating
direction method of multipliers (ADMM) optimization approach to perform compression.
Finally, the compressed QNN will be mapped to the actual quantum devices or quantum simulators for execution.

The main contributions of this paper are as follows.
\begin{itemize}
  \item \textbf{Fundamentals.} To the best of our knowledge, this is the first work to rigorously define the compression of quantum neural network, revealing the difference on the compression between quantum neural networks and classical neural networks, and the needs of compilation-aware compression.
  \item \textbf{Framework.} The paper proposes an ADMM-based holistic framework, namely \framework, to automatically compress a given quantum neural network (QNN) and deploy it to the target quantum device with a given quantum compiler.
  \item \textbf{Evaluation.} The proposed framework is evaluated on IBM Quantum processors, providing the evidence that the compression can reduce the circuit depth and in turn improve the robustness of QNNs for noisy quantum devices.  
\end{itemize}

Evaluation is conducted on both IBM Aer simulator and IBM Quantum processors.
Results on the MNIST and Fashion-MNIST dataset demonstrate that \framework can outperform vanilla VQC with significant circuit length reduction with less than 1\% accuracy degradation. Compared with existing works that consider only pruning or quantization, 
\framework can reduce the circuit depth, reaching up to $2.5\times$ with a competitive accuracy.
With 2 synthetic datasets, we further evaluate the proposed framework on IBM Quantum processors.
Compared with Vanilla VQC, \framework achieves upto 20\% accuracy improvement.
The robustness is obtained from the reduction of circuit length.



The remainder of the paper is organized as follows. Section~\ref{sec:background} reviews the related background; Section~\ref{sec:framework} presents the proposed \framework framework. Experimental results are provided in Section~\ref{Sec:Experiment} and concluding remarks are given in Section~\ref{Sec:conclusion}.

\section{Background and Related Work}
\label{sec:background}

\begin{figure}[t]
\centering
\includegraphics[width=1\linewidth]{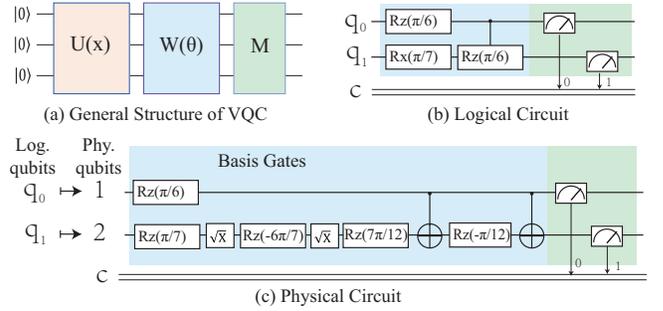}
\caption{Background of VQC. (a) A general design of VQC; (b)(c) An example of compiling virtual circuit to physical circuit through IBM Qiskit.}
\label{fig:background}
\end{figure}


\begin{figure*}[t]
\centering
\includegraphics[width=\linewidth]{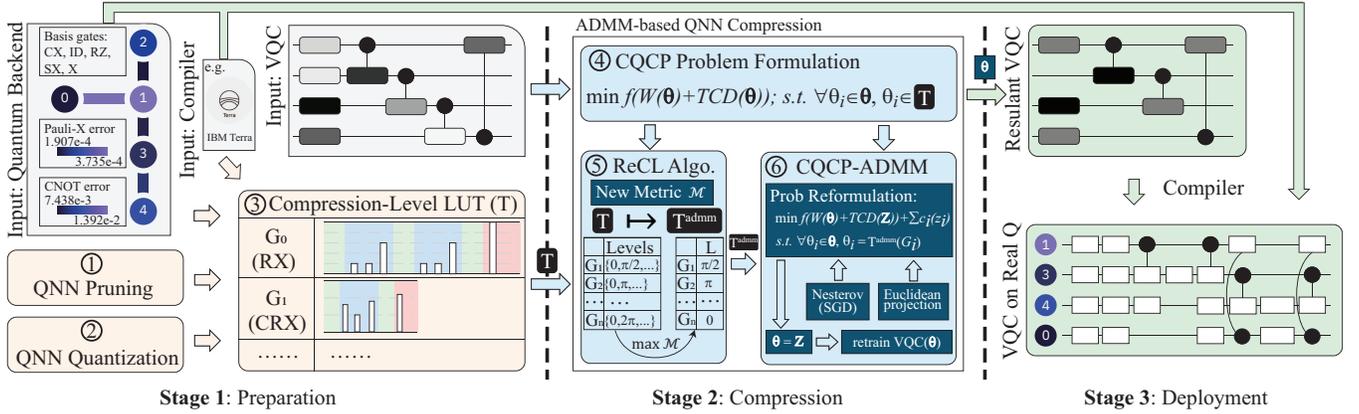}
\caption{Overview of \framework to automatically compress quantum neural networks and deploy to the actual quantum computers via 3 stages.}
\label{fig:overview}
\end{figure*}



\textbf{Variational Quantum Circuit (VQC)}. Variational quantum circuit \cite{altaisky2001quantum,ezhov2000quantum,yan2020nonlinear} is a kind of quantum neural networks (QNNs), including parameterized quantum gates, where the parameter in each gate is trainable \cite{wang2021exploration}.
Figure~\ref{fig:background}(a) illustrates a general design of a VQC, which mainly consists of three parts\cite{chen2020variational}: (1) Input Encoder $\bm{U(x)}$: quantum gates with non-adaptable parameters for encoding classical data to quantum domain; (2) Trainable Quantum Layers $\bm{W(\theta)}$: quantum gates with adaptable parameters $\bm{\theta}$; (3) Measurement $\bm{M}$:  the measurement operations to project the quantum output into the classical domain, whose output will be used for prediction.

The current VQC design~\cite{chen2020variational} is usually constructed with logical gates, namely virtual circuit, which cannot be deployed on real quantum computers directly. The virtual circuit needs to be compiled to a physical circuit at first, in order to meet the hardware constraints of the real quantum computer and then be deployed onto it. The details of the compilation are shown in Figure~\ref{fig:background}. 
More specifically, there are three steps ~\cite{venturelli2019quantum}: (1) Decompose the logical gates into physical basic gates supported by the real quantum computer; (2) Map the logical qubits to physical qubits, which produces a physical circuit; (3) Optimize the generated physical circuits for the shorter circuit depth.
After the compilation , the physical gates on the physical circuit is executed following the order of directed acyclic graph (DAG) \cite{liang2021can} of the circuit. We observed that the execution time of inference is highly coupled with the depth of the transpiled/physical circuit. Therefore, this paper utilizes transpiled circuit depth (TCD) to present the execution efficiency of VQCs.


\textbf{Deep Neural Network (DNN) Compression}. Pruning~\cite{ao2020darb,dong2020rtmobile,chang2021mix} and quantization~\cite{he2019simultaneously,gong2019differentiable,jung2019learning} are effective in reducing the model size and speed up its execution, which have been thoroughly explored for classical DNNs.
Wth the powerful ADMM optimization framework, ~\cite{zhang2018systematic} achieves a very high compression ratio while maintaining high performance. However, there are few works focusing on the application of compression techniques to QNNs.

\textbf{ADMM Basis}. 
ADMM~\cite{boyd2011distributed} is extremely capable of solving constrained optimization problems, especially those with non-convex and combinatorial constraints.
It decomposes an optimization problem to several sub-problems, solving separately.
Given a generic convex constrained optimization problem:
\begin{equation}
\min_{x} f(x)\quad \textrm{s.t.}\quad x\subseteq X
\label{eq:2}
\end{equation}
ADMM transforms this problem to an unconstrained problem, with the indicator function as follows:
\begin{equation}
I_{X}(x)=\begin{cases}
0 & \text{ if } x\in  X \\ 
+\infty  & \text{ if } x\notin  X
\end{cases}
\label{eq:3}
\end{equation}
problem(\ref{eq:2}) is transformed with an auxiliary variable $z$ and decomposes the objective function into two parts: 
\begin{equation}
\min_{x} f(x)+I_{X}(x)\quad \textrm{s.t.}\quad x=z
\label{eq:4}
\end{equation}
Finally, the Lagrangian multipliers are added to remove the constraints and the target of the optimization becomes minimizing the augmented Lagrangian function $L_\rho $:
\begin{equation}
L_{\rho}(x,z,\lambda )=f(x)+I_{X}(z)+<\lambda,x-z>+\frac{\rho }{2}\left \| x-z \right \|_{2}^{2}
\label{eq:5}
\end{equation}
where $\lambda$ is the Lagrangian multiplier and $\rho$ is a positive scalar, multiplied by a quadratic penalty term that stops $x$ and $z$ variables from being different.
The problem is solved iteratively until the solution can not be further improved.

\section{\framework Framework}~\label{sec:framework}

Figure~\ref{fig:overview} shows the overview of the proposed \framework framework. 
Given a quantum device and a quantum circuit of QNN, \framework will go through three stages to compress a QNN and deploy the compressed QNN to the quantum device. We will introduce each stage in detail in the following texts.

\vspace{5pt}
\noindent\textit{\textbf{Stage 1}. {Preparation}}
\vspace{5pt}

Stage 1 is the preparation stage. It provides the fundamentals to support the compression of QNN models.
In this stage, there are three components: \raisebox{-1pt}{\Large\ding{172}} QNN pruning, which provides the basis of what pruning is in quantum scenario; \raisebox{-1pt}{\Large\ding{173}} QNN quantization, which formulates the costs of different parameters to identify the QNN quantization; \raisebox{-1pt}{\Large\ding{174}} Compression-Level look-up-table (LUT), which is obtained based on the input quantum device and the understanding of QNN pruning and quantization.
Note that the outcome of stage 1 is the compression-level LUT, which will be the base to support the compression in stage 2.
In the following texts, we will explain each component in stage 1 in details.




\vspace{5pt}
\noindent\textbf{\raisebox{-1pt}{\Large\ding{172}} QNN Pruning: A quantum gate can be pruned not only when its parameter/weight is 0.}
\begin{equation}\label{eq:rx}
\footnotesize
RX\left(\theta\right)=\left(\begin{matrix}\cos{\frac{\theta}{2}}&-i\sin{\frac{\theta}{2}}\\-i\sin{\frac{\theta}{2}}&\cos{\frac{\theta}{2}}\\\end{matrix}\right)\ 
\end{equation}
\begin{equation}\label{eq:crx}
\footnotesize
CRX\left(\gamma\right)=\left(\begin{matrix}\begin{matrix}1&0\\0&1\\\end{matrix}&\begin{matrix}0&0\\0&0\\\end{matrix}\\\begin{matrix}0&0\\0&0\\\end{matrix}&\begin{matrix}\cos{\frac{\gamma}{2}}&-i\sin{\frac{\gamma}{2}}\\-i\sin{\frac{\gamma}{2}}&\cos{\frac{\gamma}{2}}\\\end{matrix}\\\end{matrix}\right)\ 
\end{equation}

As shown in Figure \ref{fig:cvsq}(h), in addition to the weight of 0, there exist multiple values that can make the quantum gate be pruned. And the operations corresponding to a quantum gate are the main reason of this conclusion.
Here, we make an investigation using two typical quantum gates, $RX(\theta)$ for single-qubit gates and $CRX(\gamma)$ for mult-qubit gates.
Their matrix representations are shown in Eq. \ref{eq:rx} and Eq. \ref{eq:crx}, respectively.
From these equations, we can see that parameters $\theta$ and $\gamma$ will determine the gates' function.
\begin{equation}\label{eq:ident}
\footnotesize
RX\left(0\ or\ 4\pi\right)=\left(\begin{matrix}1&0\\0&1\\\end{matrix}\right);\ CRX\left(0\ or\ 4\pi\right)=\left(\begin{matrix}\begin{matrix}1&0\\0&1\\\end{matrix}&\begin{matrix}0&0\\0&0\\\end{matrix}\\\begin{matrix}0&0\\0&0\\\end{matrix}&\begin{matrix}1&0\\0&1\\\end{matrix}\\\end{matrix}\right)
\end{equation}

When we do the computation, the quantum gates will be operated on qubits.
At each time, the system of qubits is in a specific quantum state, represented by a vector $|\psi\rangle$ (note that it can be a single-qubit system or multi-qubit system).
The state of the system is changed by the quantum gate $G$, which is represented by a unitary matrix (e.g., $G$ can be $RX(\theta)$ in Eq. \ref{eq:rx}).
Then, the computation operated on the system is a matrix-vector multiplication, i.e., $|\psi_i^\prime\rangle=G|\psi_i\rangle$.
It is clear that if $G$ is an identity matrix, then we have $|\psi_i^\prime\rangle=|\psi_i\rangle$. It indicates that the quantum states are not changed by the quantum gates, and we can prune these gates without changing the function of the quantum circuit.
In the previous examples, we observe that when the value of $\theta$ and $\gamma$ are be $0$ or $4\pi$, $RX(\theta)$ and $CRX(\gamma)$ will become the identity matrix, as shown in Eq. \ref{eq:ident}.
Kindly note that, we limit the range of parameters to $[0,4\pi]$ since other values can be mapped to this range.
\begin{equation}\label{eq:2pi}
\footnotesize
RX\left(2\pi\right)=\left(\begin{matrix}-1&0\\0&-1\\\end{matrix}\right);\ \ CRX\left(2\pi\right)=\left(\begin{matrix}\begin{matrix}1&0\\0&1\\\end{matrix}&\begin{matrix}0&0\\0&0\\\end{matrix}\\\begin{matrix}0&0\\0&0\\\end{matrix}&\begin{matrix}-1&0\\0&-1\\\end{matrix}\\\end{matrix}\right)
\end{equation}

We made another interesting observation that when $\theta$ is $2\pi$, the corresponding $RX$ gate can also be pruned. As shown in the left equation in Eq.~\ref{eq:2pi}, $RX(2\pi)$ has the negative sign for each element in the identity matrix.
Such a gate will flip the phase of each amplitude of a quantum state.
Hence, it only affects the global phase without changing the relative phases. Since the change of the global phase is not observable for the measurement, we can still prune this gate.
However, when it comes to $CRX(\gamma)$, if we set $\gamma=2\pi$, we obtain the matrix in the right-hand part of Eq. \ref{eq:2pi}.
In this case, since not all phases is flipped, we cannot prune $CRX(2\pi)$.
Based on the above observations, we give the following lemma to support the QNN pruning.
\begin{lemma}\label{lem:prune}
Given a quantum gate $G$ and parameter $\bm{\theta}$, if the function of $G(\bm{\theta})$ is to multiply the identical matrix (i.e., $I$) or the negative of identical matrix (i.e., $-I$), then $G(\bm{\theta})$ can be pruned, and $\bm{\theta}$ is thus a pruning level for $G$.
\end{lemma}

\vspace{5pt}
\noindent\textbf{\raisebox{-1pt}{\Large\ding{173}} QNN Quantization: Specified parameters/weights in quantum gates can reduce the circuit depths.}
\vspace{5pt}



In classical computing, computation using low-bit variables can reduce latency (i.e., clock cycles).
For example, compared with 32-bit floating points, the binary number (quantized from floating points) can be processed using logical gates and thus the latency is significantly reduced. We expect to improve the performance (i.e., reduce the circuit depth) of QNNs using the similar way; unfortunately, parameters in quantum computing indicate the rotation of a quantum state, and the data type will not affect the circuit depth.

However, we observe that the value of the parameter will change the circuit depth. This is because the variational quantum circuit (VQC) will finally be compiled to physical quantum processors for execution; while, the given quantum processor $\mathbb{P}$ may not directly support all logical gates in VQC and needs to compile the unsupported gates to multiple available basis gates.
For example, in Figure \ref{fig:overview}, the basis gates provided by the input quantum processor ($\mathbb{P}$) include $\mathbb{S}=\{CX, ID, RZ, SX, X\}$\footnote{For the details, please refer to \url{https://en.wikipedia.org/wiki/Quantum_logic_gate}}.
In this case, other gates (e.g., $RX$, $RY$, $U$, $CU$) have to be decomposed into the combination of these basis gates, which is contained in the compilation process.

Take $RX(\theta)$ gate as an example.
The set $\mathbb{S}$ does not contain $RX(\theta)$ gate.
To implement $RX(\theta)$ on the quantum processor $\mathbb{P}$, we compile it to realize its function using the basis gates in $\mathbb{S}$.
On IBM Quantum systems, the compiler in Qiskit (denoted as $\mathbb{C}$) will convert $RX(\theta)$ by using a sequence of $RZ$ and $SX$ gates, as below.
\begin{equation}
\footnotesize
RX\left(\theta\right)=RZ\left(\frac{5\pi}{2}\right)\cdot SX\cdot RZ\left(\theta+\pi\right)\cdot SX\cdot RZ(\frac{\pi}{2})
\end{equation}
Here, since we apply 5 sequential basis gates to express $RX(\theta)$, the circuit depth will be 5.
For special values of $\theta$, we can simplify this decomposition, and thus reduce the circuit depth.
When $\theta=3\pi/2$, we have $RX(3\pi/2)=RZ(-\pi)\cdot SX\cdot RZ(-\pi)$, where the depth is reduced to 3; and when $\theta=\pi/2$, we have $RX(\pi/2)=SX$, where the depth is further reduced to 1.

Define $len_{\mathbb{S},\mathbb{C}}(G(\theta))$ to be the function of the circuit depth of a gate $G$ (say RX) with parameter $\theta$ under the basis gate set $\mathbb{S}$ using compiler $\mathbb{C}$.
In the above $RX(\theta)$ example using IBM quantum processor with $\mathbb{S}=\{CX, ID, RZ, SX, X\}$ and IBM Qiskit's compiler $\mathbb{C}$, we have $len_{\mathbb{S},\mathbb{C}}(RX(\pi/2))=1$, and $len_{\mathbb{S},\mathbb{C}}(RX(3\pi/2))=3$.
Row ``Rx'' in Table \ref{tab:circlen} gives the circuit depth of the compiled $RX$ gate with different parameters.
Based on these definitions, we put forward the following lemma to support the QNN quantization. 

\begin{table}[t]
  \centering
  \tabcolsep 4pt
  \renewcommand\arraystretch{1.1}
  \caption{circuit depth of compiled quantum gates on IBM quantum processors; parameters are in the range of $[0,4\pi]$}
    \begin{tabular}{ccccccccccc}
    \hline
    Gate & 0     & $\pi$ & $2\pi$ & $3\pi$ & $4\pi$ & $\pi/2$ & $3\pi/2$ & $5\pi/2$ & $7\pi/2$ & others \\
    \hline
    RX    & 0     & 1     & 0     & 1     & 0     & 1     & 3     & 1     & 3     & 5 \\
    RY    & 0     & 2     & 0     & 2     & 0     & 3     & 3     & 3     & 3     & 4 \\
    CRX   & 0     & 8     & 5     & 9     & 0     & 11    & 11    & 11    & 11    & 11 \\
    CRY   & 0     & 8     & 6     & 8     & 0     & 10    & 10    & 10    & 10    & 10 \\
    \hline
    \end{tabular}%
  \label{tab:circlen}%
\end{table}%

\begin{lemma}\label{lem:quant}
Given a quantum processor with a set of basis gate $\mathbb{S}$, a quantum compiler $\mathbb{C}$, a quantum gate $G$, and a parameter $\bm{\beta}$, if $len_{\mathbb{S},\mathbb{C}}(G(\bm{\beta}))<\max_{\forall \bm{\theta}}\{len_{\mathbb{S},\mathbb{C}}(G(\bm{\theta}))\}$,
then, $G(\bm{\beta})$ is a quantized quantum gate and $\bm{\beta}$ is a quantization level for $G$ under $\mathbb{S}$ and $\mathbb{C}$.
\end{lemma}

We draw another interesting observation from Table \ref{tab:circlen}: the effects of parameters on different gates are different.
For example, with the parameter of $3\pi/2$, we can reduce the circuit depth of $RX$; however, the circuit depth cannot be reduced for $CRX$ and $CRY$.
This motivates us to build a compression-level look-up-table (LUT) for each gate, as discussed below.





\vspace{5pt}
\noindent\textbf{\raisebox{-1pt}{\Large\ding{174}} Compression-Level LUT: Different quantum gates have varied pruning and quantization levels.}
\vspace{5pt}

\begin{figure}[t]
\centering
\includegraphics[width=1\linewidth]{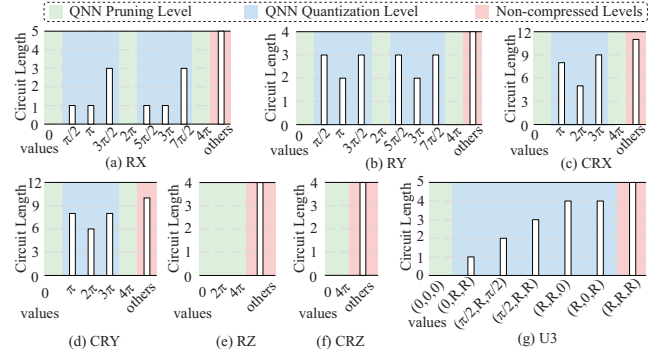}
\caption{Compression-level LUT, including different quantum gates involved in the input VQC, based on the basis gate set $\mathbb{S}=\{CX,ID,RZ,SX,X\}$ and IBM Qiskit compiler.}
\label{fig:LUT}
\end{figure}

As shown in Table \ref{tab:circlen} and Figure ~\ref{fig:LUT}, not only quantization levels but also pruning levels of quantum gates are different.
We uniformly call them ``compression level'' if the level can do either pruning or quantization.
Here, one compression level may not be a single parameter, but a tuple $\langle\theta,\phi,\lambda\rangle$.
This depends on the parameterized quantum gate.
For example, $U3$ and $CU3$ gates have three Euler angles/parameters.

Given a set of basis gate $\mathbb{S}$ and a compiler $\mathbb{C}$, component \noindent\textbf{\raisebox{-1pt}{\Large\ding{174}}} in \textit{Stage 1} will generate a compression-level LUT for all gates involved in the input VQC.
As an example, Figure \ref{fig:LUT} illustrates the LUT built by \framework.
Kindly note that, since $\mathbb{S}$ contains $RZ$ gate, Figures \ref{fig:LUT}(e)-(f) only contain the pruning levels without quantization levels.
For the ease of illustration, we only present the parameters within the range of $[0,\pi/2]$ of $U3$ gates in Figure \ref{fig:LUT}(g).

Based on Lemma \ref{lem:prune}, Lemma \ref{lem:quant}, and the compression-level LUT built by \framework, we have the following Corollary in \textit{Stage 1}.
\begin{corollary}
The quantum neural network (QNN) compression is to reduce the depth of the compiled quantum circuit, which is affected by the set of basis gates supported by the given quantum device and the quantum compiler. That is, QNN compression needs compilation awareness.
\end{corollary}

\vspace{5pt}
\noindent\textit{\textbf{Stage 2}. {Compression}}
\vspace{5pt}

Before introducing the details of how to compress QNNs, we first formulate the compiler-aware QNN compression problem below.

\vspace{5pt}
\noindent\textbf{\raisebox{-1pt}{\Large\ding{175}} Compilation-aware QNN Compression Problem (CQCP).}
\vspace{5pt}

As shown in Figure \ref{fig:overview}, there are two inputs of stage 2: (1) a parameterized VQC circuit, denoted as $W(\bm{\theta})$, where $\bm{\theta}$ is the trainable parameters and $\theta_i\in\bm{\theta}$ is the parameter associated with the $i^{th}$ quantum gate ($G_i$); (2) the compression-level LUT, denoted as $T$. 
For the gate $G_i$, $T(G_i)$ is a set, including all potential compression levels of $G_i$.
For example, if $G_0$ is a $CRY$ gate, then we have $T(G_0)=\{0, \pi, 2\pi, 3\pi, 4\pi\}$ (see Figure \ref{fig:LUT}(d)); (3) a user-specified quantum compiler $\mathbb{C}$.



The CQCP problem is defined as follows: Given a VQC $W(\bm{\theta})$, a compression-level LUT $T$, and a quantum compiler $\mathbb{C}$, the problem is to determine the trainable parameters $\bm{\theta}$, such that 
\begin{equation}
\begin{aligned}
  &\min_{\{\bm{\theta}\}} \quad f(W(\bm{\theta})) +  TCD(\bm{\theta})  \\
  &s.t. \quad\quad\forall \theta_i \in\bm{\theta}, \theta_i \in T(G_i)
\end{aligned}
\label{eq:general_train}
\end{equation}
where function $f$ represents training loss on the given dataset, and function $TCD(\bm{\theta})$ is the circuit depth of VQC $W(\bm{\theta})$ compiled by the compiler $\mathbb{C}$.

In the regular QNN training, we have $T(G_i)\in \mathbb{R}$ (i.e., no constraint), and thus the optimal gates ($ G_{i}(\theta_{i})$) for each layer can be obtained by classical gradient-based methods, e.g. stochastic gradient descent (SGD).
However, if we directly integrate the QNN compression in the training process, the constraint of $T(G_i)$ becomes combinatorial and non-convex, which prevents the Eq (\ref{eq:general_train}) from being solved through gradient-based methods.
Besides, $TCD(\bm{\theta})$ can only be obtained from the quantum circuit compilation, which is not a derivable function.


To solve the aforementioned constrained non-convex optimization CQCP problem, there are two possible directions: (1) directly apply the existing mathematical programming techniques \cite{fang2018spider,chi2019nonconvex}, such as employing a stochastic path-integrated differential estimator; (2) transform the constrained non-convex problem to an unconstrained convex problem, such as employing alternating direction method of multipliers (ADMM) approach \cite{boyd2011distributed}.
The large search space of CQCP brings challenges for both directions, that is, 
the number of possible combinations of all parameters (s.t., $\theta_i \in T(G_i)$) exponentially grows along with the number of gates in VQC, implying an exponential number of constraints (called ``constraint set'').
According to previous work~\cite{srinivasan1992ritual}, the constraint set is highly degenerate, which tends to cause standard nonlinear program solvers, i.e., direction (1), getting bogged down in many steps without convergence of the cost function.
In this paper, we explore the solution on direction (2).

\vspace{5pt}
\noindent\textbf{\raisebox{-1pt}{\Large\ding{176}} Compression-Level LUT Reconstruction for ADMM}
\vspace{5pt}

To enable ADMM approach, the first and foremost challenge is how to deal with the large search space, while not affecting the model performance.
Our design principle is to reconstruct the constraints of CQCP problem in two ways: (1) we downsize the compression-level LUT, i.e., $T$ $\mapsto$ $T^{admm}$, aiming at reducing the size of search space; (2) we relax the constraints that all parameters have to be compressed, i.e., $\forall \theta_i\in\bm{\theta}, \theta_i\in T(G_i)$ by enabling a portion of parameters to be real numbers to improve model performance.
In this way, we expect that we can reduce the search space without corrupting the model performance.

In component \raisebox{-1pt}{\Large\ding{176}}, we will present an algorithm, namely $ReCL$, to reconstruct the compression-level LUT: $T$ $\mapsto$ $T^{admm}$.
More specifically, we will select the most appropriate compression level for each gate as the constraint but not a list of compression levels, i.e., $\forall G_i$, $T^{admm}(G_i)\in T(G_i)$.
The downsizing process is conducted based on a new metric, $\mathcal{M}$,
which can quantify the impact of the gate value on accuracy and TCD simultaneously. 
Given a VQC $W(\bm{\theta})$, the metric $\mathcal{M}(\bm{\theta}, G_i(\gamma_{i,k}))$ describes the impact of $G_i$ by changing $\theta_i\in\bm{\theta}$ to $\gamma_{i,k}$ (i.e., the $k^{th}$ compression level in $T(G_i)$).
Let $\bm{\theta^{i,k}}$ be the new parameters, where $\forall \theta_j\in \bm{\theta}, j\ne i$, we have $\theta^{i,k}_j=\theta_j$, and we have $\theta^{i,k}_i=\gamma_{i,k}$.
Then, we have the below formula:
\begin{equation}
\begin{aligned}
\mathcal{M}(\bm{\theta},G_i(\gamma_{i,k})) = acc(W(\bm{\theta^{i,k}})) \cdot \tau(\bm{\theta^{i,k}},\bm{\theta}) 
\label{eq:metric}
\end{aligned}
\end{equation}
where $acc(W(\bm{\theta^{i,k}}))$ is the accuracy VQC under the new parameters, and $\tau(\bm{\theta^{i,k}},\bm{\theta})$ indicates the compression ratio by changing parameters from $\bm{\theta}$ to $\bm{\theta^{i,k}}$, that is, $\tau(\bm{\theta^{i,k}},\bm{\theta}) = \frac{TCD(\bm{\theta^{i,k}})}{TCD(\bm{\theta})}$.
In this way, we can make sure accuracy and TCD value in the 
same numerical standard.

On top of the new metric, $ReCL$ algorithm is to reconstruct the compression-level LUT ($T^{admm}$) by setting $T^{admm}(G_i)$ to $\gamma_{i,k}$ yield to the maximum $\mathcal{M}(\bm{\theta},G_i(\gamma_{i,k}))$.
More specifically, the process is conducted by traversing all quantum gate in VQC, and iteratively change the parameter of one gate $G_i$ in the following two steps.
\begin{itemize}[noitemsep,topsep=0pt,parsep=0pt,partopsep=0pt]
    \item Traverse all compression level $\gamma_{i,k}\in T(G_i)$ and 
    obtain metric $\mathcal{M}(\bm{\theta},G_i(\gamma_{i,k}))$.
    \item Set $T^{admm}_{ij}$ be $\gamma_{i,k}$ which leads to the highest metric, i.e. $max_{\forall \gamma_{i,k}\in T(G_i)}[\mathcal{M}(\bm{\theta},G_i(\gamma_{i,k}))]$.

\end{itemize}

From these steps, we can reconstruct and downsize compression-level LUT, which reduces the constraint set of the ADMM and accelerates the training convergence.



\vspace{5pt}
\noindent\textbf{\raisebox{-1pt}{\Large\ding{177}} ADMM-Based Compilation-aware QNN Compression}
\vspace{5pt}

Based on the reconstructed compression-level LUT $T^{admm}$, we will decompose the original CQCP problem into subproblems that can be solved separately and iteratively until convergence using ADMM, denoted as CQCP-ADMM.
As mentioned in component \raisebox{-1pt}{\Large\ding{176}}, in ADMM optimization, we relax the constraint to enable a portion of parameters to be not compressed in each iteration: $T^{admm}\mapsto T^{s,r}$, where $T^{s}$ is the relaxed compression constraint set, and $r$ is the index of iterations.

With these understandings, we reformulate the CQCP (Eq. \ref{eq:general_train}),
\begin{equation}
\begin{aligned}
  &\min_{\{ \theta_i\}} \quad f(W(\bm{\theta})) +TCD(\bm{Z})+ \sum_{\forall z_i\in\bm{Z}}{c_i(Z_i)},  \\
  &s.t. \quad\quad\quad \forall \theta_i \in\bm{\theta}, \quad \theta_i = T^{admm}(G_i).
\end{aligned}
\label{eq:admm_train}
\end{equation}
where $\bm{Z}$ is a set of auxiliary variables for subproblem decomposition and $z_i\in \bm{Z}$ is corresponding to $\theta_i\in \bm{\theta}$; $f(W(\bm{\theta}))+TCD(\bm{Z})$ corresponds to the objective function in the original CQCP problem (Eq.  \ref{eq:general_train}).
Besides, we use an indicator function (Eq. \ref{eq:indicator_func}) to move the constraint set toward the objective function in each iteration $r$.
\begin{equation}
\begin{aligned}
c_i(Z_i)=\begin{cases}
 0& \text{ if } \theta_i \in T^{s,r}(G_i), T^{s,r} = T^{admm}\odot mask^r,\\ 
 +\infty & \text{ if } otherwise. 
\end{cases}
\end{aligned}
\label{eq:indicator_func}
\end{equation}
where $mask^r$ is built to indicate the parameters will be compressed at iteration $r$.
In detail, $mask^r$ is built by three steps:
\begin{enumerate}
\item $\forall G_i$, it calculates the $distance^r_i$ between $\theta_i^{r+1}+\lambda^r$ and $T^{admm}(G_i)$ and the transpiled gate $depth$, $LUT[T^{admm}(G_i)]$;
\item We sort the importance of all gates in terms of $\alpha$$\cdot$$distance^r_i$$+$$(1-\alpha)$$\cdot$$depth$, ($0<\alpha<1$);
\item According to the required compression rate $\tau$, $mask^r(G_i)=1$ if $distance^r_i$ is in the front of $\tau\times |\bm{G}|$ gates, otherwise $mask^r(G_i)=0$. Notation $|\bm{G}|$ is the total number of gates.
\end{enumerate}
With $mask^r$, we define operation $T^{admm}\odot mask^{r}$ to relax the condition of compression on gate $G_i$ and formulate $T^{s,r}$; in particular, we have $T^{s,r}(G_i)=T^{admm}(G_i)$, if $mask^{r}=1$.

By reformulating CQCP problem, it is decomposed into three subproblems in Eq. \ref{eq:admm_train}, which can be solved according to the augmented Lagrangian function~\cite{fortin2000augmented}.
\begin{equation}
\small
\begin{aligned}
  &\theta_{i}^{r+1} = {\arg\min}_{\theta_i} f(W(\theta_{i}))+\sum_{i=1}^N \frac{\rho_i}{2}||\theta_i-Z_i^r+\rho_i^{-1}\lambda _i^r||^2_2, \label{eq:admm_sub1}
\end{aligned}
\end{equation}
\begin{subequations}
\small
\begin{align}
 &Z_i^{r+1} = {\arg\min}_{Z_i} \sum_{i=1}^N [\frac{\rho_i}{2}||\theta_i^{r}-Z_i+\rho_i^{-1}\lambda _i^r||^2_2 + c_{i}(Z_i)]+TCD(Z),\label{eq:admm_sub2:obj}\\
 &\textrm{s.t.}\quad Generate\ mask^r\  by\  step (1)\sim(2)\ of mask\ building\ process,                                                   \label{eq:admm_sub2:c1}\\
 &Z_i^{r+1} = 
\begin{cases}
Z_i^{r} , & \textrm{otherwise}, \\
LUT_{s}(\theta_{i}^{r+1}+\lambda_i^{r}), & \textrm{if } G_i \in mask^r.
\end{cases} \label{eq:admm_sub2:c2}
\end{align}
\label{eq:admm_sub2}
\end{subequations}
\begin{equation}
\small
\begin{aligned}
&\lambda^{r+1}=\lambda^{r} + \rho(\theta^{r+1}-Z^{r+1}). \label{eq:admm_sub3}
\end{aligned}
\end{equation}

The first sub-problem (\ref{eq:admm_sub1}) can be solved using a general gradient-based method or Nesterov’s algorithm.
Note that sub-problem (\ref{eq:admm_sub1}) is convex.
It minimizes the training loss similar to (\ref{eq:general_train}), while it tends to produce $\theta_i$ close to $Z_i$  in order to minimize the second term in the objective function as well. $\rho$ is the hyperparameters for the regularization factor of the second term.

Solving the second sub-problem (\ref{eq:admm_sub2}) ensures that the gate parameters $\bm{\theta}$ corresponding to $\bm{Z}$ satisfies the compression-level constraints, $\theta_i$$\in$$T_i^{s}$, and minimizes the TCD. 
Essentially, optimizing the sub-problem (\ref{eq:admm_sub2}) produces a $\bm{Z}$ with minimal changes to the $\theta^{r}+\rho^{-1}\lambda^{r}$ vector and projects this vector into the feasible region of gate parameters. 
More specifically, $TCD(\bm{Z})$ is a function of the transpiled depth of the different gates (Table~\ref{tab:circlen},Fig ~\ref{fig:LUT}).
And the solution of $Z_i^{r+1}$ is obtained by (\ref{eq:admm_sub2:c2}) which is the Euclidean projection, i.e. $LUT_s$ the look-up-table operation of $T_i^{s,r}$, and avoids the derivation of $c(\bm{Z})$ and $TCD(\bm{Z})$.
Please note that a single iteration of CQCP-ADMM produces a solution ($\bm{Z}$) that meets the $T^{s}$ constraints, while the result may be sub-optimal in terms of the training loss. 
Hence, multiple iterations of the CQCP-ADMM algorithm provide a high-quality solution on both the training loss and the TCD reduction.

The third sub-problem (\ref{eq:admm_sub3}) updates the $\lambda$ with the generated $\theta$ and $Z$ to accelerate the convergence of the CQCP-ADMM.
If $\theta_i^{r+1}$ and the corresponding $Z_i^{r+1}$ are close to each other, then $\lambda_i^{r+1}$ will be close to $\lambda_i^{r}$.
Otherwise, by updating $\lambda_i^{r+1}$, the divergence of $\theta_i$ and $Z_i$ values will be minimized in the next iteration of the CQCP-ADMM.
The stopping metric for the CQCP-ADMM algorithm is defined as follows,
\begin{equation}
||\theta_i^{r}-\theta_i^{r+1}||_2^2< \zeta,\quad ||Z_i^{r}-Z_i^{r+1}||_2^2< \zeta
\label{eq:stopping_metric}
\end{equation}
Finally, based on the gap between $\theta$ and $Z$ solutions, one can terminate the algorithm once the gap is lower than a certain threshold.
Due to the small size of the VQC at this stage, in our experience, 10$\sim$15 iterations of the CQCP-ADMM algorithm are enough to generate superior solutions both on the accuracy performance and TCD reduction of the VQC.

In the final step of the CQCP-ADMM algorithm, we choose $Z$ as the final solution, and the compression requirements and the optimal TCD reduction is met, yet the solution may be sub-optimal in terms of the accuracy performance of the VQC.
At this step, we generate the compression position (mask) from the $Z$ and retrain the VQC with the mask to recover the accuracy performance, which can ensure the maximum TDP compression with a slight accuracy degradation.

\vspace{5pt}
\noindent\textit{\textbf{Stage 3}. {Deployment}}
\vspace{5pt}

After we obtain the compressed VQC from Stage 2, \framework will leverage the input compiler to do the transpilation that maps the logical quantum circuit to the given quantum device.






\section{\tool: Experiment}\label{Sec:Experiment}

\subsection{Experiment Setups}

\noindent\textbf{Datasets.} We evaluate \framework on 2 common classification datasets (MNIST~\cite{lecun1998gradient} and Fashion MNIST~\cite{xiao2017fashion}) and one synthetic dataset.
For MNIST, we extract 2 classes (i.e., digits 3 and 6), denoted as ``MNIST-2''.
For Fashion-MNIST, we extract 2-4 classes from dress, shirt, and T-shirt/top, denoted as ``Fashion-MNIST-2'', ``Fashion-MNIST-3'', ``Fashion-MNIST-4'', respectively.
All data are downsampled to $4\times 4$ dimensions.
We generate two synthetic datasets, denoted as ``Syn-Dataset-4'' and ``Syn-Dataset-16'' to indicate the data with 4 features (i.e., a 4-dimension input vector) and 16 features, respectively.
We generate 100 data with 2 classes ($C1$ and $C2$) using 2 normal distributions ($D1$ and $D2$).
For class $C1$, the front half of features (i.e., front 2 features for Syn-Dataset-4) follows $D1$, while the tail half of features follows $D2$; data in class $C2$ are generated on the opposite way.
For all datasets, we apply 90\% samples for the train set and 10\% for test set.







\noindent\textbf{QNN models.}
According to the dimension $N$ of inputs, the QNN involves $\log_2(N)$ qubits.
The circuit contains 3 parts: encoding, computation, and measurement.


\textit{Encoding.}
We apply `amplitude encoding' presented in \cite{jiang2021co} for Fashion-MNIST and angle encoding' presented in \cite{larose2020robust} for MNIST and synthetic datasets.
For amplitude encoding, we convert the input values to the amplitudes by L2-normalization.
For angle encoding, each input dimension is associated with 1 rotation gate and the value is encoded to the phase of the associated rotation gate.
For 16-dim input, we use 4RY, 4RZ, 4RX, and 4RY gates; while for 4-dim input, we use 2RY and 2RZ gates.



\textit{Computation.}
All models consist of `RX', `RY', `RZ', `CRX', `CRY', and `CRZ' gates. Each gate has one trainable parameter.
We apply the VQC designs in \cite{chen2020variational} to build QNN for different datasets. 
There are 50, 50, 30, 22, 22, and 14 for MNIST-2, Fashion-MNIST-4, Fashion-MNIST-3, Fashion-MNIST-2, Syn-Dataset-16, and Syn-Dataset-4, respectively.

\textit{Measurement.}
The classification results are obtained based on the measurement results, which is the expectation values of states on Pauli-Z basis. 
Note that for 3 classes on 4 qubits, we divide the first 15 states into 3 groups and sum up amplitudes in each group to generate the output values.
Then we process the output values by Softmax to get probabilities for classification.


\noindent\textbf{Quantum devices and compiler configurations.}
We use both IBM Qiskit Aer simulator and IBM-Q quantum computers via Qiskit APIs. 
Specifically, we apply Qiskit Terra as the compiler for all QNN circuits.  
The real-quantum device executions are conducted on ``ibm\_largos'', ``ibm\_perth'', or ``ibm\_jakarta'' backends. 
 
\noindent\textbf{Competitors.}
We compare the proposed \framework against three competitors: (1) Zero-Only Pruning \cite{wang2021quantumnas}, which prunes the VQC model without considering Compilation; (2) CompVQC-Pruning, which is based on \framework by only involving pruning levels in the compression-level LUT; (3) CompVQC-Quant, which is also based on \framework and only involve quantization levels in the compression-level LUT.
For all experiments, we use the ``Vanilla VQC'' without compression as the baseline.

\subsection{Results on Qiskit Aer}

\begin{table}[]
\small
\centering
\tabcolsep 1.5 pt
\renewcommand{\arraystretch}{1.1}
\caption{Comparison among different methods on the accuracy performance and the TCD of the VQC}
\label{tab:main result-simulation}
\begin{tabular}{|c|cc|cc|}
\hline
\multirow{3}{*}{\textbf{\begin{tabular}[c]{@{}c@{}}Compression \\ Method\end{tabular}}} & \multicolumn{2}{c|}{\textbf{MNIST-2}}                                                                                    & \multicolumn{2}{c|}{\textbf{Fashion-MNIST-2}}                                                                            \\ \cline{2-5} 
                                                                                        & \begin{tabular}[c]{@{}c@{}}Acc.\\ (vs. Baseline)\end{tabular} & \begin{tabular}[c]{@{}c@{}}TCD\\ (Speedup)\end{tabular} & \begin{tabular}[c]{@{}c@{}}Acc.\\ (vs. Baseline)\end{tabular} & \begin{tabular}[c]{@{}c@{}}TCD\\ (Speedup)\end{tabular} \\ \hline
Vanilla VQC                                                                            & 82.74\%(0)                                                   & 121(0)                                                    & 87.58\%(0)                                                   & 92(0)                                                     \\ \hline
Zero-Only-Pruning                                                                       & 80.58\%(-2.16\%)                                               & 70(1.73$\times$)                             & 86.92\%(-0.67\%)                                               & 63(1.46$\times$)                                                  \\ \hline
CompVQC-Pruning                                                                         & 81.83\%(-0.91\%)                                               & 74(1.64 $\times$)                             &87.41\%(-0.17\%)                                                & 47(1.96$\times$)                                                  \\ \hline
CompVQC-Quant                                                                   & 80.99\%(-1.75\%)                                                & 108(1.10$\times$)                            & 86.25\%(-1.33\%)                                               & 74(1.24$\times$)                                                  \\ \hline
CompVQC                                                                                 & \textbf{81.83\%(-0.91\%)}                                               & \textbf{47(2.57$\times$)}                             & \textbf{87.58\%(-0.00\%)}                                                  & \textbf{47(1.96$\times$)}                                                 \\ \hline
\end{tabular}
\end{table}

\noindent\textbf{Comparison among compression approaches.}
Table \ref{tab:main result-simulation} reports the comparison results of different compression methods on MNIST-2 and Fashion-MNIST-2.
We compare both accuracy and circuit depth (TCD), using Vanilla VQC as a baseline.
Since TCD indicates the latency of the quantum circuit, the reduction in TCD can be expressed by the speedup achieved by the corresponding methods.

From Table~\ref{tab:main result-simulation}, on MNIST-2, our proposed \framework performs best among all competitors.
Specifically, compared with Vanilla VQC, \framework has the lowest TCD, achieving $2.57\times$ reduction with a negligible accuracy drop ($-0.91\%$).
On the other end, \framework-Quant performed poor on both metrics, with a large accuracy loss (-1.75\%) and a slight TCD reduction (1.10$\times$).
It is not difficult to understand this result that pruning removes entire gates with a similar loss of accuracy, but quantization only reduces the number of physical quantum gates.
On this basis, we conclude that quantum pruning is a more effective compression type than quantum quantization.

Another interesting observation is made on two pruning methods: \framework-pruning and Zero-Only-Pruning.
Both methods can achieve 1.6$\times$-1.7$\times$ reduction on TCD, but Zero-Only-Pruning suffers a severe drop in accuracy performance ($2.16\%\downarrow$); while \framework-pruning achieves the same accuracy as \framework.
When comparing the pruning-only methods to \framework, we observe that \framework can achieve significant TCD reduction nearly without accuracy drop.


All the above results show that only prune parameters approaching to 0 cannot provide the best performance, and we need to involve compilation/traspilation in the compression process to generate accurate and efficient quantum circuits; furthermore, with both pruning and quantization, it will lead to the best performance.





Similarly, the results on Fashion-MNIST-2 in Table~\ref{tab:main result-simulation} draw the similar conclusion.
The only exception is \framework and \framework-Pruning can consistently reduce TCD.
Nevertheless, the accuracy of \framework is still a little higher than \framework-Pruning.
These results inspired us to explore the tradeoff between accuracy and TCD.


\begin{figure}[t]
\centering
\includegraphics[width=1\linewidth]{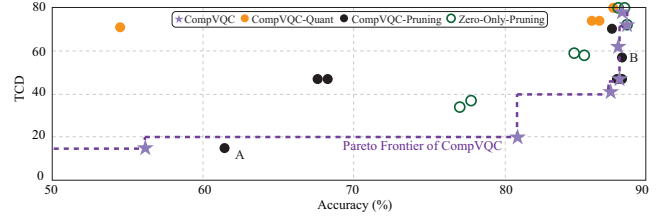}
\caption{Main results: The Accuracy-Circuit Depth Tradeoff on Fashion-MNIST2}
\label{fig:depth-acc}
\end{figure}

\noindent\textbf{Accuracy vs. TCD.}
Figure ~\ref{fig:depth-acc} explores the accuracy-TCD tradeoff for different approaches on Fashion-MNIST2 dataset.
The x-axis and y-axis represent accuracy and TCD respectively.
With the objective in Eq. \ref{eq:general_train}, the right-bottom corner will be the ideal solution, indicating the highest accuracy and the lowest TCD.
In the figure, the dotted line represents the Pareto frontier obtained by our proposed \framework.
It is clear that this Pareto frontier dominates all solutions of \framework-Quant and Zero-Only-Pruning.
In addition, for \framework-Pruning, there are only two solutions escaping the Pareto frontier of \framework (point A and point B in the figure).
These results emphasize that the proposed \framework can make the best tradeoff between accuracy and quantum circuit depth.

\begin{figure}[t]
\centering
\includegraphics[width=3.1018 in]{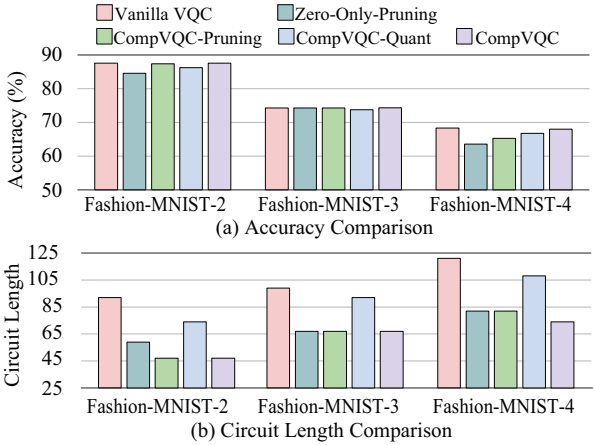}
\caption{Main Results: \framework Scalability on Fashion-MNIST with 2-4 class }
\label{fig:fashion-234}
\end{figure}
\noindent\textbf{Scalability on different datasets.}
Figure \ref{fig:fashion-234} reports the results of all compression approaches on Fashion-MNIST datasets with 2-4 classes.
Figure \ref{fig:fashion-234}(a) shows that \framework can achieve higher accuracy over other competitors, which is the closest one to the vanilla VQC.
From Figure \ref{fig:fashion-234}(b), we can see that pruning or quantization methods can significantly reduce the circuit depth (TCD), and \framework can consistently obtain the lowest TCD for all datasets.
Since the dataset with more classes has more quantum gates, the above results show that the proposed approach can consistently work for different scales of input VQC circuits.
\begin{table*}[h]
\tabcolsep 2pt
\small
\centering
\renewcommand{\arraystretch}{1.3}
\caption{Optimize different VQC architectures with different numebr of layers on Fashion-MNIST-2 dataset by \framework}
\label{tab:different architecture}
\begin{tabular}{|c|c|cccc|cccc|cccc|cccc|}
\hline
\multicolumn{2}{|c|}{\multirow{2}{*}{Method}} & \multicolumn{4}{c|}{QLayer22}                                                                & \multicolumn{4}{c|}{QLayer30}                                                                & \multicolumn{4}{c|}{QLayer38}                                                                & \multicolumn{4}{c|}{QLayer50}                                                                \\
                                \multicolumn{2}{|c|}{} & \multicolumn{1}{c}{Acc.} &    vs. BL.                         & TCD & vs. BL.                          & \multicolumn{1}{c}{Acc.} &   vs. BL.                          & TCD &   vs. BL.                        & \multicolumn{1}{c}{Acc.} &    vs. BL.                         & TCD &    vs. BL.                       & \multicolumn{1}{c}{Acc.} &     vs. BL.                        & TCD &  vs. BL.                         \\ \hline
\multicolumn{2}{|c|}{Vanilla VQC}                    & 87.58\%                      &                  baseline           & 92  &                   baseline        & 87.58\%                      &      baseline                       & 99  &      baseline                     & 90.17\% &       baseline                                            & 117 &     baseline                      & 90.33\%                      &     baseline                        & 121 &        baseline                   \\ \hline
CompVQC & (Accuracy)                & 88.08\%                      & \multicolumn{1}{c}{0.50\%}  & 72  & \multicolumn{1}{c|}{$1.28\times$} & 88.08\%                      & \multicolumn{1}{c}{0.50\%}  & 72  & \multicolumn{1}{c|}{$1.38\times$} & 90.08\%                      & \multicolumn{1}{c}{-0.08\%} & 103 & \multicolumn{1}{c|}{$1.14\times$} & 89.25\%                      & \multicolumn{1}{c}{-1.08\%} & 78  & \multicolumn{1}{c|}{$1.55\times$} \\ \hline
CompVQC & (TCD)                     & 87.00\%                      & \multicolumn{1}{c}{-0.58\%} & 41  & \multicolumn{1}{c|}{$2.24\times$} & 87.00\%                      & \multicolumn{1}{c}{-0.58\%} & 42  & \multicolumn{1}{c|}{$2.36\times$} & 87.33\%                      & \multicolumn{1}{c}{-2.83\%} & 38  & \multicolumn{1}{c|}{$3.08\times$} & 87.75\%                      & \multicolumn{1}{c}{-2.58\%} & 51  & \multicolumn{1}{c|}{$2.37\times$} \\ \hline
\end{tabular}
\end{table*}
\subsection{Results on IBM Quantum Processors}

Table~\ref{tab:main result-real qc} reports the results on both Qiskit Aer simulator and the actual IBM Quantum processors using two synthetic datasets.
As shown in this table, \framework can reduce circuit depth by 2.09$\times$ and $2.22\times$ on Syn-Dataset-4 and Syn-Dataset-16, respectively.
On accuracy, it is clear that \framework can maintain the accuracy to be the same on the noisy real quantum computers for both datasets; on the other hand, the Vanilla VQC has a 15\% and 10\% accuracy drop on two datasets respectively, after deploying the VQC to the real quantum processors.
The above results demonstrate that the 
quantum compression can reduce circuit depth, and in turn make the QNN more robust to the noise on the real quantum processors.
This also points out that quantum compression facilitates the practical deployment of quantum circuits on real machines.

We further conduct a group of experiments with Syn-Dataset-4 on different backends of IBM Quantum processors.
Results are reported in Table \ref{tab:different backend}.
It is clear that for 3 backends (i.e., ``ibm\_lagos'', ``ibm\_perth'', and ``ibm\_jakarta'') with different noise levels, the propsoed \framework can always achieve over 98\% accuracy, while the accuracy of Vanilla VQC is from 79\% to 92\%.
Results show that \framework obtain robust VQC target different quantum devices.

\begin{table}[t]
\small
\centering
\tabcolsep 1.5 pt
\renewcommand{\arraystretch}{1.1}
\caption{Case studies on IBM Quantum Processors}
\label{tab:main result-real qc}
\resizebox{\linewidth}{!}{
\begin{tabular}{|cc|cccc|}
\hline
\multicolumn{2}{|c|}{\textbf{Datasets}}                                                    & \multicolumn{2}{c}{\textbf{Syn-Dataset-4}}                                                                                                                 & \multicolumn{2}{c|}{\textbf{Syn-Dataset-16}}                                                                                                               \\ \hline
\multicolumn{2}{|c|}{}                                                                    &                                                                                &                                                                              &                                                                                &                                                                              \\
\multicolumn{2}{|c|}{\multirow{-2}{*}{\textbf{Compression Method}}}                       & \multirow{-2}{*}{\begin{tabular}[c]{@{}c@{}}Acc.\\ (vs. Baseline)\end{tabular}} & \multirow{-2}{*}{\begin{tabular}[c]{@{}c@{}}TCD\\ (Speedup)\end{tabular}} & \multirow{-2}{*}{\begin{tabular}[c]{@{}c@{}}Acc.\\ (vs. Baseline)\end{tabular}} & \multirow{-2}{*}{\begin{tabular}[c]{@{}c@{}}TCD\\ (Speedup)\end{tabular}} \\ \hline
 \multicolumn{1}{|c|}{\multirow{2}{*}{Qiskit Aer}}                & Vanilla VQC & 94\%(0)                                                                        & 23(0)                                                                        & 96\%(0)                                                                        & 51(0)                                                                        \\ \cline{2-6} 
\multicolumn{1}{|c|}{}                        & Comp-VQC     & 99\%(5\%)                                                                     & 11($2.09\times$)                                                             & 98\%(2\%)                                                                     & 23($2.22\times$)                                                             \\ \hline
\multicolumn{1}{|c|}{}                             & Vanilla VQC & 79\%(-15\%)                                                                     & 23($1.00\times$)                                                             & 86\%(-10\%)                                                                     & 51($1.00\times$)                                                             \\ \cline{2-6} 
\multicolumn{1}{|c|}{\multirow{-2}{*}{IBM Q}} & Comp-VQC     & 99\%(5\%)                                                                     & 11($2.09\times$)                                                             & 98\%(2\%)                                                                     & 23($2.22\times$)                                                             \\ \hline
\end{tabular}
}
\end{table}

\begin{table}[t]
\small
\centering
\tabcolsep 7 pt
\renewcommand{\arraystretch}{1.1}
\caption{\framework on different backends}
\label{tab:different backend}
\begin{tabular}{|c|ccc|}
\hline
Acc.(vs. Baseline) & ibm\_lagos & ibm\_perth & ibm\_jakarta \\ \hline
Vanilla VQC(TCD=23)     & 79\%(0)    & 86\%(0)    & 92\%(0)       \\ \hline
CompVQC(TCD=11)          & 99\%(20\%) & 98\%(12\%) & 100\%(8\%)    \\ \hline
\end{tabular}
\end{table}
\subsection{Ablation Studies}

\begin{figure}[t]
\centering
\includegraphics[width=1\linewidth]{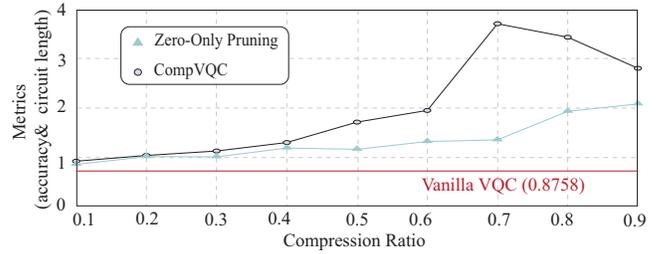}
\caption{\framework on different compression ratio}
\label{fig:ratio-metrics}
\end{figure}

All the above results have already demonstrated the superiors of \framework and the scalability of \framework on different datasets.
In this subsection, we conduct ablation studies on different compression ratio and different sizes of VQC on the same dataset.

\noindent\textbf{A. \framework consistently outperforms compilation-agnostic compression at different Compression Ratios} 

Figure \ref{fig:ratio-metrics} reports the comparison of \framework and Zero-Only-Pruning on different ratios.
The y-axis indicates the metric in Eq. \ref{eq:metric}, which is the higher the better.
As a reference, the metric of vanilla VQC is 0.875.
From this figure, it is clear that \framework consistently achieves higher metric over the existing Zero-Only-Pruning; moreover, the improvement is enlarged along with the increase in compression ratio from 0.1 to 0.7. When the compression ratio reaches 0.8, we observe a decrease of metrics on \framework, this is mainly because the accuracy is decreased.
Overall, we can see that \framework consistently performs better than the existing compilation-agnostic compression.


\noindent\textbf{B. On the same dataset, \framework can significantly reduce circuit depth with guaranteed accuracy.}

Table \ref{tab:different architecture} reports the results of \framework on Fashion-MNIST-2 dataset applying VQC with a different number of gates. Note VQC will repeat a block/layer by multiple times to integrate more parameters.
As shown in this table, with more gates, say ``QLayer50'' with 50 gates, Vanilla VQC can obtain higher accuracy.
In the table, we have two rows of \framework, the one with ``(Accuracy)'' indicates the results with the superior accuracy in the exploration results, while the other one with ``(TCD)'' is for the exploration with an accuracy threshold $87.00\%$.
From this table, it is clear that the accuracy-oriented \framework can reduce circuit depth while maintaining high accuracy, for the VQC with 22 and 30 layers, \framework can achieve 1.28$\times$ and 1.38$\times$ reduction on circuit depth, while obtaining 0.5\% accuracy gain.
In addition, for the TCD-oriented \framework, it can achieve more than $2\times$ reduction on circuit depth while maintaining the accuracy to be at least 87\%.
These results again show the scalability of the proposed framework and its ability to make the best accuracy and circuit depth tradeoff.

\vspace{-5pt}
\section{Conclusion}\label{Sec:conclusion}
\vspace{-5pt}
In this work, we first study the fundamental of the compression (including both pruning and quantization) in QNN, revealing the difference of compression between QNNs and classical DNNs.
We proposed \framework, a framework that automatically compresses a given quantum neural network and deploys it to the target quantum device with a specific quantum compiler.
Evaluations on MNIST \& Fashion-MNIST show the effectiveness of \framework, which is the superior among the three competitors.
We also deploy the vanilla model and compressed model by \framework to real IBM quantum devices, and find that \framework can make the QNN more robust to the noise by reducing the circuit depth. 
We envision the fundamental of QNN compression and the optimization studied in this paper will push the real machine learning applications to the near-term quantum devices.

\bibliographystyle{ACM-Reference-Format}
\bibliography{ref}

\end{document}